\documentclass{article}\usepackage{natbib,emulateapj}
\usepackage{psfig}
\newcommand{\beq}{\begin{equation}}
\newcommand{\eeq}{\end{equation}}
\newcommand{\bea}{\begin{eqnarray}}
\newcommand{\eea}{\end{eqnarray}}
\newcommand{\bean}{\begin{eqnarray*}}
\newcommand{\eean}{\end{eqnarray*}}
\newcommand{\ba}{\begin{array}}
\newcommand{\ea}{\end{array}}
\newcommand{\bml}{\begin{mathletters}}
\newcommand{\eml}{\end{mathletters}}
\newcommand{\dt}[1]{\frac{\partial{#1}}{\partial t}}
\newcommand{\dx}[1]{\frac{\partial{#1}}{\partial x}}
\bibpunct[,]{(}{)}{;}{a}{}{,}
\begin{document}
\title{Thermal Conduction and the Stability of Hot Accretion Flows}
\author{Mikhail V. Medvedev$^{1,2,}$\altaffilmark{3} and Ramesh
Narayan$^2$} \affil{$^1$ Canadian Institute for Theoretical
Astrophysics, University of Toronto, Toronto, ON, M5S 3H8, Canada }
\affil{$^2$ Harvard-Smithsonian Center for Astrophysics, 60 Garden
Street, Cambridge, MA 02138 }
\altaffiltext{3}{medvedev@cita.utoronto.ca;
http://www.cita.utoronto.ca/$\sim$medvedev/}

\begin{abstract}
Recently, Medvedev \& Narayan (2001) discovered a new type of
accretion flow, a hot settling flow around a rapidly rotating neutron
star.  The flow is cooling-dominated and energetically similar to the
Shapiro, Lightman, \& Eardley (1976, SLE) solution.  Since the SLE
solution is known to be thermally unstable, one might suspect that the
new solution would also be unstable.  However, due to the very high
temperature of the accreting gas, thermal conduction is very strong
and could suppress the thermal instability.  We analyze the role of
thermal conduction in both the hot settling flow and the SLE solution.
In the hot settling flow collisions are very rare.  Therefore, thermal
transport occurs via free streaming of electrons along tangled
magnetic field lines.  We find that conduction is strong enough to
make the flow marginally stable.  In contrast, in the cooler SLE
solution, conduction is via collisional, Spitzer-type transport.  In
this case, conduction is weaker, and we find that the SLE solution is
thermally unstable even in the presence of conduction.
\end{abstract}

\section{Introduction}

Accretion flows around compact objects frequently radiate significant
levels of hard X-rays, indicating the presence of hot optically-thin
gas in these systems.  A number of hot accretion solutions have been
discussed in the literature, e.g., the \citet{SLE76} (SLE) solution,
the advection-dominated accretion flow (ADAF)
\citep{NY94,NY95a,NY95b,Abram+95}, and the convection-dominated
accretion flow (CDAF) \citep{NIA00,QG00}.  These solutions are
relevant for accretion onto a black hole.  Recently, \citet{MN00}
discovered a solution that corresponds to accretion onto a neutron
star (NS).  We refer to this solution as a ``hot settling flow,''
since the gas is hot (the temperature is nearly virial) and it
``settles'' onto the NS (the ratio of the radial velocity to the
free-fall velocity decreases with decreasing radius).

Not all hot accretion flows are stable.  The cooling-dominated SLE
solution has been shown to be thermally unstable
\citep{P78,WL91,NY95b} and, hence, unlikely to exist in nature.  More
generally, it has been shown that any accretion flow in which heating
balances cooling is thermally unstable if the cooling is due to
bremsstrahlung emission \citep{SS76,P78}. The ADAF solution, on the
other hand, is known to be thermally stable
\citep{NY95b,K+96,K+97}. In this solution, cooling is weak (ideally
zero), and so the thermal energy of the flow is not radiated but is
advected with the gas (hence the name). The CDAF is also believed to
be stable, since in this flow again the thermal energy is advected by
convective eddies and is either carried into the black hole or is
radiated near the outer boundary of the flow \citep{BNQ01}.

The thermal stability of the \citet{MN00} hot settling flow has not
been addressed so far.  Energetically, this flow is very similar to
the SLE solution, since the heat energy produced by viscous
dissipation is radiated locally via bremsstrahlung.  One might
therefore expect the flow to be thermally unstable.  However, this is
not necessarily the case, as we show in this paper.

The temperature of the gas in all these accretion solutions is very
high, reaching almost the virial temperature in several cases.
Thermal conduction is thus likely to be enormous and could have a
significant effect.  The question that we try to answer in this paper
is: How does conduction affect the thermal stability of the hot
settling flow and the SLE solution?

In the hot settling flow, the mean-free-path is much larger than the
flow scale (the radius); therefore, the Spitzer theory
\citep{Spitzer62} of thermal conduction cannot be applied.  Instead we
need to apply the transport theory for collisionless plasmas with
tangled (but dynamically unimportant) magnetic fields
\citep{RR78,CC98,MK00,Mal00}.  In the cooler SLE solution, collisions
are more frequent; hence it is appropriate to use collisional
transport theory modified for the presence of tangled fields.
  
The paper is organized as follows. We begin with a formal discussion
of the thermal instability in \S\ref{S:TI}, and discuss the
stabilizing effect of thermal conduction in \S\ref{S:STAB}. In
\S\ref{S:SETTL} and \S\ref{S:SLE} we study the effect of thermal
conduction on the hot settling flow and the SLE flow, respectively,
and we conclude with a discussion in \S\ref{S:CONCL}.

\section{Thermal Instability in an Accretion Flow}
\label{S:TI}

The physics of the thermal instability is simple \citep{F65}.  Suppose
a system is in thermal equilibrium, so that the rates of heating and
cooling per unit volume are equal: $Q^+=Q^-$.  For simplicity let us
take the heating and cooling rates to be functions of only the local
temperature: $Q^+\propto T^\alpha$, $Q^-\propto T^\beta$ ($\alpha,\
\beta>0$ for concreteness).

Suppose, with increasing temperature, the cooling rate rises faster
than the heating rate, i.e., $\beta>\alpha$. Then a local perturbation
which causes a small increase in the temperature will result in a net
cooling of the gas: $Q^->Q^+$. This will cause the temperature to
return to its equilibrium value, which means that the gas will be
thermally stable.  (It is easily seen that this is true also for a
small decrease in the temperature.)  On the other hand, if
$\alpha>\beta$, the gas is thermally unstable. For instance, if the
temperature decreases slightly, cooling becomes stronger than heating
and the system deviates from its equilibrium in a run-away manner.

To study the thermal stability of an accretion flow, we need to
include additional physics, namely the effects of shear and rotation.
The shearing sheet approximation \citep{GLb65,GT78} is a convenient
way of introducing the relevant physics without unnecessary technical
complications.  This approximation is quite accurate for perturbations
on length scales much smaller than the local radius.

Conventionally, the shearing sheet coordinates are Cartesian with $x,\
y,\ z$ corresponding to the radial, azimuthal, and vertical
directions, respectively. These coordinates are appropriate for
describing the motion of a parcel of gas whose geometrical size is
small compared to the local radius, $R$, of the flow (i.e., $x,\ y,\
z\ll R$), so that the effects of geometry and curvature are
insignificant. It is convenient to compare the wave-vector $k$ of a
perturbation with $1/R$ and the frequency of a mode with the local
Keplerian frequency $\Omega_K=\sqrt{GM/R^3}$, where $M$ is the mass of
the central object.  The shearing sheet approximation is accurate for
``local'' small-scale perturbations with $kR\gg1$. Perturbations with
$kR\sim1$ are global; their properties may be understood only through
a global stability analysis.

We consider a shearing gas flow with unperturbed velocity given by
\beq {\bf V}_0(x)=2A\,x\,\hat{y}, \eeq where $2A=d{\bf V}_0/dx$ is the
shear frequency and ``hat'' denotes a unit vector.  Note that we have
neglected the radial velocity in the equilibrium flow since this
component of the velocity is significantly smaller than the azimuthal
velocity in both the hot settling flow and the SLE solution.  To
include the effect of rotation we assume that there is a Coriolis
acceleration, described by an angular rotation frequency ${\bf
\Omega}=\Omega\,\hat{z}$. The vorticity and epicyclic frequency are
then given by \beq 2B=2A+2\Omega,\qquad \kappa_{\rm epi}=2(\Omega
B)^{1/2}.  \eeq Both the hot settling flow and the SLE solution
satisfy the Keplerian scaling, $\Omega\propto R^{-3/2}$.  Therefore,
for both solutions we have $2A=-(3/2)\Omega$, $2B=\Omega/2$ and
$\kappa_{\rm epi}=\Omega$.

We assume that perturbations in the flow have structure only in the
$x$ direction, and we ignore motions in the $z$ direction.  We write
the perturbations (represented by primes) in the velocity, density and
sound speed as \bean {\bf
V}'(x,t)&=&u(x,t)\,\hat{x}+v(x,t)\,\hat{y},\\
\rho'(x,t)&=&\rho_0\sigma(x,t), \\ {c_s^2}'(x,t)&=&a^2(x,t), \eean
where $\rho_0$ and $c_s^2$ are the equilibrium values of the density
and the square of the sound speed.  Note that we define $c_s$ to be
the isothermal sound speed, so that the pressure is written as $p=\rho
c_s^2$.  By considering perturbations of the basic hydrodynamic
equations, namely the continuity, radial momentum, azimuthal momentum
and entropy equations, we obtain the following four equations,
\bml\bea & &\dt{\sigma}+\dx{u}=0,\\ & &\dt{u}-2\Omega
v+c_s^2\dx{\sigma}+\dx{a^2}=0,\\ & &\dt{v}+2Bu=0,\\ &
&\frac{\rho_0}{\gamma-1}\dt{a^2}-\rho_0 c_s^2\dt{\sigma}
=\left(Q^++Q^-\right)'\label{energy}, \eea\label{eqs}\eml where we
have used $d/dt=\partial/\partial t+V_{0x}\partial/\partial x\simeq
\partial/\partial t$ since the inflow velocity $V_{0x}$ is set to zero
in our approximation.  Note that gravity does not enter the perturbed
equations (it of course enters the unperturbed equation, where in
equilibrium it cancels the centrifugal acceleration).  In the shearing
sheet approximation, the gravitational acceleration is assumed to be
independent of $x$; therefore, it does not contribute to the perturbed
equations.  For simplicity, we have neglected viscosity in the
azimuthal momentum equation, though we do include viscous dissipation
in the energy equation through the terms $Q^+$ and $Q^{+'}$.

For the heating and cooling rates, we make use of ``realistic'' expressions 
that represent the physics of viscous accretion flows. Thus we write 
\beq 
Q^+=\alpha\frac{\rho c_s^2}{\Omega_K} \left(\frac{d
V_{0y}}{dx}\right)^2=4\alpha A^2\frac{\rho c_s^2}{\Omega_K}, \qquad
Q^-=-{\cal C}\rho^2\left(c_s^2\right)^n,
\label{heating}
\eeq 
where $\alpha\sim0.1$ is the standard Shakura-Sunyaev viscosity parameter,
$V_{0y}$ is the $y$-component of the unperturbed velocity, and ${\cal C}$ 
is a constant.  We leave the index $n$ in the cooling function
unspecified for now, but we note that $n=1/2$ corresponds to
non-relativistic free-free (bremsstrahlung) cooling.  In equilibrium,
we have $Q^+_0+Q^-_0=0$, and for the perturbations we find
$$
(Q^++Q^-)'=-4A^2\alpha\,\frac{\rho_0c_s^2}{\Omega_K}
\left(\sigma+(n-1)\,\frac{a^2}{c_s^2}-\frac{1}{A}\frac{dv}{dx}\right).
$$ 

We assume that the perturbations in equations (\ref{eqs}) are of the
form $\exp(-i\omega t+ikx)$.  Substituting in the above equations and
solving, we obtain the following dispersion relation: \bea \lefteqn{
\omega\left[\frac{\omega}{\gamma-1}+\frac{i(n-1)}{\tau_{\rm
cool}}\right] \left(\omega^2-\kappa_{\rm epi}^2-k^2c_s^2\right){} }
\nonumber\\ & &{}\qquad\qquad
-\omega\left[\omega+\frac{i(2B/A-1)}{\tau_{\rm
cool}}\right]k^2c_s^2=0,
\label{disp0}
\eea where \beq \tau_{\rm cool}=\left(
\frac{\rho_0c_s^2}{Q_0^\mp}\right)
=\frac{\Omega_K}{4A^2\alpha}=\frac{4}{9\alpha s^2}\,\Omega_K^{-1}
\label{tau-cool}
\eeq is the cooling (heating) time of the gas and $s=\Omega/\Omega_K$
is the dimensionless angular velocity of the gas.  

The dispersion relation (\ref{disp0}) corresponds to purely radial
perturbations.  The same relation can be used also for perturbations
in the vertical direction, except that we must set $\kappa_{\rm
epi}=0$.  Perturbations in the azimuthal direction are more
complicated.  Because of the shear, a non-axisymmetric wave packet is
distorted as a function of time, and must be analyzed by special
techniques which are beyond the scope of this paper (see, e.g.,
\citealp{T77,GT78}).

Equation (\ref{disp0}) is a fourth-order polynomial and has four roots
corresponding to four modes.  A flow is unstable if any of the four
modes grows with time, i.e. if the corresponding root has ${\rm
Im}\;\omega>0$.  One of the roots of the dispersion relation is always
$\omega=0$.  This root corresponds to the viscous mode, which in the
present case is particularly simple because we neglected viscosity in
the momentum equation.  It is easy to show that if we introduce
viscosity into the momentum equation the viscous mode would become
stable, i.e., we will obtain ${\rm Im}\;\omega<0$.  We do not consider
the viscous mode further in this paper.

The physics of the remaining three modes may be understood by
considering equation (\ref{disp0}) in various limits.  Consider first
the limit $k\to0$.  In this limit, two of the roots are given by
$\omega=\pm\kappa_{\rm epi}$, corresponding to simple epicyclic
oscillations.  In the opposite limit $k\to\infty$, the same roots are
given by $\omega=\pm \gamma^{1/2}c_sk$, which shows that they
correspond to sound waves.  In the absence of heating and cooling
(i.e. $\tau_{\rm cool}\to\infty$), we can obtain an exact solution for
these roots which is valid for all $k$: \beq \omega^2=\kappa_{\rm
epi}^2+{\gamma}c_s^2k^2.\label{dispacoust} \eeq This is the standard
dispersion relation for sound waves in a differentially rotating flow.
The presence of $\gamma$ is because the relevant sound speed is the
adiabatic sound speed, $\gamma^{1/2}c_s$ (recall that $c_s$ is defined
to be the isothermal sound speed).

The final root of the dispersion relation (\ref{disp0}) corresponds to
the thermal mode.  In the limit $k\to0$, we obtain \bml\beq
\omega=i\,(\gamma-1)\,\frac{(1-n)}{\tau_{\rm cool}}.  \eeq We see that
the mode is stable (for $\gamma>1$) if $n>1$ and unstable if $n<1$.
In the opposite limit $k\to\infty$, we find \beq
\omega=i\,\frac{(\gamma-1)}{\gamma}\,\frac{(2-n-2B/A)}{\tau_{\rm cool}},
\label{omega0}
\eeq\eml Now, the mode is stable if $n>2(1-B/A)$, i.e. $n>8/3$ for our
problem, and unstable otherwise.  Note that an accretion flow that is
cooled by free-free emission ($n=1/2$) is unstable in both limits.

The dashed lines in Fig. \ref{f:disp} show the real and imaginary
parts of the various roots of the dispersion relation (\ref{disp0}) as
functions of $k$ for a realistic set of parameters. We have chosen
$\kappa_{\rm epi}^2 =0.5\Omega_K^2$ which corresponds to a spin
parameter%
\footnote{The value of $s\sim0.7$ was chosen such as to illustrate the effect 
	of rotation which is proportional to $\kappa_{\rm epi}^2\propto s^2$. 
	This value of $s$ is somewhat higher than a typical spin of rotating 
	neutron stars, which is around $s\sim0.1$. For such $s$ the 
	effect of rotation is negligible.}
 $s=\sqrt{0.5}$, and we have set the free-free cooling time
to be $\tau_{\rm cool,ff}=10/\Omega_K$ which corresponds to
$\alpha\simeq0.1$ (see equation [\ref{tau-cool}]).  Since we are
interested in hot accretion flows with nearly virial temperature, we
expect the sound speed to be comparable to the free-fall velocity;
therefore, we have set $c_s= \Omega_K R$.  Other parameters are
$\gamma=5/3$ and $n=1/2$.

In Fig. \ref{f:disp} we do not show the trivial viscous root
$\omega=0$.  We label the two acoustic modes (propagating radially in
opposite directions) as 1 and 2 and we label the thermal mode as
3. The frequencies of the acoustic modes are modified from the
analytic solution given in equation (\ref{dispacoust}), which was
derived by neglecting cooling.  The real parts of the roots are
slightly perturbed, and both roots pick up an imaginary part.
However, the imaginary parts are negative, which means that these
modes are stable in the presence of cooling. The thermal mode,
however, is unstable (${\rm Im}\;\omega>0$) for all $k$
(Fig. \ref{f:disp}), as expected from the asymptotic analysis
presented earlier.  We focus on this mode in the rest of the paper.

\section{Stabilizing Effect of Thermal Conduction}
\label{S:STAB}

It is easy to see that thermal conduction will tend to reduce the
thermal instability.  An unstable thermal mode of wave-vector $k$
consists of a growing temperature perturbation of wave-length
$2\pi/k$.  Thermal conduction tends to smooth out this temperature
perturbation through heat diffusion.  If the rate at which the
temperature perturbation grows is smaller than the rate at which it is
smoothed out by conduction, then the instability will be suppressed
and the mode will be stable. Otherwise, the mode will continue to
grow, but at a somewhat reduced rate.  

The rate at which fluctuations are smoothed out by conduction depends
on the spatial scale of the perturbation. The smaller the scale
(i.e. the larger the value of $k$), the faster the conduction, and the
greater the stabilizing effect.  Thus, we expect conduction to
stabilize thermal modes with $k$ greater than some critical $k_{\rm
crit}$.  Our task in this section and succeeding sections is to
estimate $k_{\rm crit}$ through a quantitative analysis.  If we find
that $k_{\rm crit}R\gg1$, then we conclude that the flow is thermally
unstable.  On the other hand, if we find that $k_{\rm crit}R \lesssim
1$, we may reasonably claim that the flow is thermally stable.
Technically, for $k\sim1/R$, we need to carry out a global analysis
rather than the local analysis presented in this paper, but this is
beyond the scope of the present paper.

Let us write  the heat flux $q$ due to thermal conduction as
\beq q_{\rm cond}=-\kappa\nabla T,
\label{q-grad}
\eeq
where $\kappa$ is the thermal conductivity coefficient. Thermal 
conductivity in a dense, fully ionized gas is given by the
\citet{Spitzer62} formula,
\beq
\kappa_{\rm Sp}\approx1.3nk_Bv_T\lambda 
\simeq6.2\times10^{-7}T_e^{5/2}\textrm{ erg/(s K cm)}.
\label{kappa-Sp}
\eeq Here $v_T=(k_BT_e/m_e)^{1/2}$ is the electron thermal speed,
$T_e$ is the electron temperature ($T_e=T$ for a one-temperature
plasma), $k_B$ is the Boltzmann constant, and \beq \lambda\simeq10^4
T_e^2/n\textrm{ cm}
\label{mfp}
\eeq is the electron mean free path.  Note that $\lambda$ is
independent of the mass of the particle.

In the collisionless regime, i.e., when the mean free path of an
electron becomes comparable to or larger than the temperature gradient
scale $\lambda\ga T_e/|\nabla T_e|$, equation (\ref{q-grad}) for the
heat flux is no longer valid.  For an unmagnetized plasma, the heat
flux takes the following saturated form \citep{CMc77}, \beq q_{\rm
sat}\simeq-C\rho c_s^3\textrm{ sgn}(\nabla T),
\label{q-sat}
\eeq where $C\sim5$ is a numerical constant whose exact value depends
on the particle distribution function.  This result is not relevant
for our problem since our plasma is magnetized.

For a collisionless magnetized plasma, thermal conduction is
anisotropic.  Electrons stream freely along the field lines, and the
parallel heat flux remains the same as for the unmagnetized case
described above.  However, the transverse heat flux is greatly reduced
because electrons are tied to the field lines on the scale of the
Larmor orbit.  In fact, if the field is uniform and homogeneous, the
perpendicular thermal flux is identically equal to zero since
electrons cannot move across the field lines.  In a tangled field,
however, electrons can jump from one field line to another and thus
conduct heat perpendicular to the field.  Since we are dealing with a
turbulent accretion flow with a tangled magnetic field, this is the
regime of interest to us.

The physics of this regime of conduction has been discussed by
\citet{RR78} and \citet{CC98}, who identified two important effects
which are discussed in more detail in Appendix \ref{A:COND}. 
 
First, since particles can move freely only along field lines, the
characteristic effective mean free path is set by the correlation
scale of the magnetic field $l_B$.  In a hot accretion flow this scale
is not known in general. However, it is likely that turbulent motions
in the flow occur on a scale comparable to the local radius $R$, since
this is the only characteristic scale in the problem.  Very likely,
the turbulent magnetic field will also have the same scale $l_B\sim
R$.  We parameterize this scale as $l_B=\xi R$.  We expect $\xi\le1$
because turbulent fluctuations cannot have a scale larger than the
local radius of the flow. We assume $\xi\sim0.1$ throughout the paper.

Second, the magnetic field is inhomogeneous.  Therefore, only a
fraction $\vartheta<1$ of the particles will be able to pass though
the magnetic mirrors that will be present in the field, and it is only
these particles that transport energy beyond a distance $\sim l_B$.
For magnetic field strength fluctuations $\delta B\sim\langle
B\rangle$, the fraction of free streaming particles is estimated to be
$\vartheta\sim0.3$.

Typically, hot accretion flows are highly collisionless, i.e.,
$\lambda\gg R\ga l_B$.  Therefore, we can write the thermal conduction
coefficient as \beq \kappa_B\simeq nk_Bv_Tl_B\,\vartheta \simeq
10^{-2}nk_Bv_TR\xi_{-1}\vartheta_{-1},
\label{kappa_B-am}
\eeq where $\xi_{-1}=\xi/10^{-1}$ and
$\vartheta_{-1}=\vartheta/10^{-1}$.  Let us write the conductive heat
flux in a form similar to that used for the viscous stress, namely
\beq q_{\rm cond}=-\alpha_c\frac{c_s^2}{\Omega_K}\rho\frac{dc_s^2}{dx},
\label{q-cond}
\eeq where the dimensionless coefficient $\alpha_c$ is analogous to
the Shakura-Sunyaev viscosity parameter $\alpha$, and is given by \beq
\alpha_c \simeq\frac{R}{H}\,\xi\vartheta\, F(e,p)
\simeq10^{-2}\xi_{-1}\vartheta_{-1}\, F(e,p).
\label{alpha_c}
\eeq Here we have used the fact that $v_T\simeq c_{se}$ and $H/R\sim
c_s/v_{\rm ff} \sim c_s/\Omega_KR$, where $H$ is the accretion disk
scale height (in hot flows, $H\sim R$) and $v_{\rm ff}$ is the
free-fall speed.  The quantity $F(e,p)$ takes into account whether the
conduction is dominated by protons or electrons; its numerical value
is given in equation (\ref{F}).

In the presence of conduction the energy equation has an additional
contribution from the divergence of $q_{\rm cond}$.  The perturbed
energy equation (3d) then becomes modified to \bea
\frac{\rho_0}{\gamma-1}\dt{a^2}-\rho_0 c_s^2\dt{\sigma}=(Q^+-Q^-)'
+\alpha_c\frac{\rho c_s^2}{\Omega_K}\frac{\partial^2 a^2}{\partial
x^2}.
\eea This equation together with equations (3a-c) yields the
following modified dispersion relation: \bea \lefteqn{
\omega\left[\frac{\omega}{\gamma-1}+\frac{i(n-1)}{\tau_{\rm cool}}
+\frac{ik^2R^2}{\tau_{\rm cond}}\right] \left(\omega^2-\kappa_{\rm
epi}^2-k^2c_s^2\right){} } \nonumber\\ & &{}\qquad\qquad\quad
-\omega\left[\omega+\frac{i(2B/A-1)}{\tau_{\rm
cool}}\right]k^2c_s^2=0,
\label{disp1}
\eea
where 
\beq
\tau_{\rm cond}=\Omega_KR^2/\alpha_c c_s^2
\label{tau-cond}
\eeq is the conductive time scale. 

Figure \ref{f:disp} illustrates how thermal conduction modifies the
thermal instability. The solid lines in the two panels show the real
and imaginary parts of the roots of the dispersion relation
(\ref{disp1}) as functions of $k$. The parameters have the same values
as before, namely $\kappa_{\rm epi}/\Omega_K=s=\sqrt{0.5}$,
$c_s/\Omega_K R=1$, $\tau_{\rm cool,ff}\Omega_K=10$, $\gamma=5/3$,
$n=1/2$.  We have set the conductive time equal to $\tau_{\rm
cool,ff}$ (for illustration), which corresponds to
$\alpha_c\sim0.3$. 

Figure \ref{f:disp} shows that the imaginary parts of all three modes
decrease rapidly with increasing $k$.  For the particular parameters
we have selected, the growth rate of the unstable thermal mode (curve
3) goes to zero at $k_{\rm crit}R\sim1.5$, and the mode is stable for
all $k>k_{\rm crit}$.

We may also analyze equation (\ref{disp1}) analytically. In the
large-$k$ limit, the root corresponding to the thermal mode is equal
to\footnote{ In deriving this equation from (\ref{disp1}) we have used
the fact that the acoustic time-scale is, in general, shorter than the
time-scale of the thermal mode, i.e., $\omega\ll k c_s$, and we have
neglected $\kappa_{\rm epi}$ as before. In this case we can neglect
$\omega^2$ in the second brackets, so that equation (\ref{omega1})
readily follows.  It may seem that this procedure fails when the
acoustic and thermal time-scales are comparable. This may happen when
$kR\sim1$ (for such perturbations, the sound crossing time is of order
the dynamical time) and when $\tau_{\rm cool}$ and $\tau_{\rm cond}$
are also comparable to the dynamical time, which is
$\sim\Omega_K^{-1}$. Nevertheless, even in this case, equation
(\ref{omega1}) works fairly well near the stability threshold. Indeed,
at the threshold itelf, ${\rm Im}\ \omega=0$. Since further the
thermal mode frequency has no real part, we have $\omega\sim0$ near
the threshold; so we may safely neglect $\omega^2$ compared to
$k^2c_s^2$ in the second brackets of equation (\ref{omega1}).}  \beq
\omega= i\,\frac{(\gamma-1)}{\gamma}\left(\frac{(2-n-2B/A)}{\tau_{\rm
cool}} -\frac{k^2R^2}{\tau_{\rm cond}}\right).
\label{omega1}
\eeq Clearly, for large $k$, conduction stabilizes the thermal mode,
for the reasons explained at the beginning of this section.  Using the
above relation, we can estimate the critical $k_{\rm crit}$ above
which all $k$ are stable: \beq k_{\rm crit}^2R^2=\frac{\tau_{\rm
cond}}{\tau_{\rm cool}}\left(2-n-2\,\frac{B}{A}\right)
=\frac{13}{6}\,\frac{\tau_{\rm cond}}{\tau_{\rm cool}},
\label{stab-T}
\eeq where we have substituted $n=1/2$ (free-free cooling) and
$B/A=-1/3$ (Keplerian scaling).

We should comment here that in the theory described above, the
conductivity $\kappa_B$ is a quantity averaged over many field
correlation lengths.  Therefore, the results of the stability analysis
are valid only for perturbations on scales much larger than $l_B$.
(This is somewhat inconsistent since we have assumed that $l_B=\xi
R\sim0.1R$.)  For small-scale perturbations with $k\gg (\xi R)^{-1}$,
the local magnetic field is nearly homogeneous. Therefore, thermal
conductivity is anisotropic; its perpendicular component is of order
of $\kappa_B$ (equation [\ref{kappa_B}]), while the conductivity along
the field lines is much larger:
$$ \frac{q_\|}{q_\bot}=\frac{C\rho c_s^3\textrm{ sgn}(\nabla T)}
{-\alpha_c(\rho c_s^2/\Omega_K)(dc_s^2/dx)} \sim \frac{C\rho
c_s^3}{\alpha_c\rho c_s^3(H/R)} \sim\frac{5}{\alpha_c}\gg1, $$ as
follows from equations (\ref{q-cond}) and (\ref{q-sat}).  Thermal
instability along the magnetic field lines is then suppressed more
strongly than in the analysis presented above.

\section{Thermal Stability of the Hot Settling Flow Solution}
\label{S:SETTL}

The hot settling flow solution \citep{MN00} describes an optically
thin two-temperature accretion flow onto a rotating neutron star.  The
flow extracts the rotational energy of the star and radiates it via
free-free emission.  The main properties of the solution are
summarized in Appendix \ref{A:SETTL}.  

Because the hot settling flow is cooling-dominated and radiates by
free-free emission, it is intrinsically thermally unstable, as follows
from equation (\ref{omega0}) for $\gamma>1,\ n=1/2$, and
$B/A=-1/3$. However, thermal conductivity in the flow is enormous
because of the very high temperature and large mean free path of
particles.  The coefficient of thermal conduction is estimated in
Appendix \ref{A:SETTL} to be \beq
\alpha_c\simeq10^{-2}\xi_{-1}\vartheta_{-1}F(e,p), \eeq where $F(e,p)$
lies in the range $1\le F(e,p)\la15$.  Is this level of thermal
conduction enough to stabilize the thermal mode?

The relevant stability criterion is given in equation (\ref{stab-T}).
However, before we apply this criterion, we need to allow for the fact
that, as shown in Appendix \ref{A:SETTL}, thermal conduction in the
flow is so strong that it modifies even the equilibrium structure of
the flow.  In particular, the cooling time (\ref{tau-cool}) is
modified as per the substitution given in equation (\ref{sub}) and
becomes \beq \tau_{\rm cool} =\frac{4}{(9\alpha
s^2+2\alpha_c)}\,\Omega_K^{-1}
\simeq\frac{2}{\alpha_c}\,\Omega_K^{-1}, \eeq where we have assumed
that $\alpha_c\gg\alpha s^2$, which is reasonable for typical
parameters, e.g. $\alpha\sim0.1,\ s\sim0.1$, $\alpha_c\sim0.1$.  From
equation (\ref{tau-cond}) and using the self-similar solution
(\ref{ss}), we obtain the thermal conductive time\footnote{
Alternatively, we recall that in the hot settling flow $H/R\sim 1$
and $H/R\simeq c_s/v_{\rm ff}\simeq c_s/(\sqrt2 \Omega_K R)$. Then
$\tau_{\rm cond}$ readily follows from equation (\ref{tau-cond}).}
\beq \tau_{\rm cond}=\frac{3}{\alpha_c}\,\Omega_K^{-1}, \eeq 
Substituting $\tau_{\rm cool}$ and $\tau_{\rm cond}$ into the
stability criterion (\ref{stab-T}), we then find \beq k_{\rm
crit}R=\left[\frac{26\alpha_c}{9(9\alpha s^2+2\alpha_c)}\right]^{1/2}
\simeq\sqrt{\frac{13}{9}}\simeq1.2,
\label{kR-stab-ss}
\eeq that is, thermal modes with $kR\ga 1$ are stable.  This suggests
that the hot settling flow with thermal conduction is marginally
stable to the thermal instability.

Whether the mode $kR=1$ itself is stable or not cannot be reliably
determined from our local analysis.  A global stability analysis is
necessary to properly account for the effects of geometry and
curvature, but this is beyond the scope of the present paper.

\section{Thermal Stability of the SLE solution}
\label{S:SLE}

The SLE solution \citep{SLE76} was the first hot optically thin
accretion solution discovered.  This self-similar solution is best
suited for accretion onto a black hole, though it could in principle
be applied also for accretion onto a NS, by attaching a suitable
boundary layer at its inner edge.  The SLE solution as originally
envisaged by Shapiro et al. (1976) was cooled by Compton-scattering of
soft photons from an outer thin accretion disk.  A local form of the
solution (e.g., \citealp{NY95b}) involves local cooling via free-free
emission.  Some properties of this solution are discussed in Appendix
\ref{A:SLE}.  The solution is thermally unstable \citep{P78}, as
discussed in \S2.  The question we consider here is whether thermal
conduction in the SLE solution is strong enough to eliminate the
instability.

In Appendix \ref{A:SLE} we show that the SLE flow is cooler than the
hot settling flow and hence more collisional. Therefore, the
collisionless thermal conductivity given by equation (\ref{kappa_B})
is not applicable.  One should use instead equation
(\ref{kappa_B-SLE}).  To decide whether thermal conduction has any
hope of stabilizing the mode, we determine an upper bound on the
coefficient of thermal conduction.  From equations (\ref{kappa_B-SLE})
and (\ref{kappa_Sp-SLE}) we find \beq
\kappa_{SLE}\la 1.1\times10^{20}\vartheta \dot
m^{5/4}\alpha^{-5/2}r^{-15/8}.
\label{kappa_SLE}
\eeq The cooling time then follows from (\ref{tau-cool}) for a
Keplerian flow ($s=1$): \beq \tau_{\rm cool}=\frac{\rho
c_s^2}{Q^-}=\frac{8}{9\alpha}\Omega_K^{-1}.
\label{SLE-tau}
\eeq Comparing equations (\ref{q-grad}) and (\ref{q-cond}), we write
$\kappa=\alpha_c\rho c_s^2k_B/(\Omega_Km_p)$.  We then obtain a lower
limit on the conductive time from equations (\ref{tau-cond}) and
(\ref{kappa_SLE}): 
\bea \tau_{\rm cond}&=&\frac{\Omega_KR^2}{\alpha_cc_s^2} 
=\left(\frac{k_B}{m_p}\frac{\rho c_s^2}{\kappa_{SLE}\Omega_K}
\frac{R^2}{H^2}\right)\Omega_K^{-1} \nonumber \\
&\ga&13.3\vartheta^{-1}\dot m^{-1}\alpha^{-2}r^{1/2}\Omega_K^{-1}.
\label{tau-cond-SLE}
\eea Here we have used the values of plasma parameters for a
two-temperature zone (\ref{SLE-sol}), which yields a shorter
$\tau_{\rm cond}$ than for a one-temperature zone.  We notice that the
conduction time is significantly longer than the cooling time for
typical parameters of the flow, e.g., $\dot m\sim0.01,\
\alpha\sim0.1,\ \vartheta\sim0.1,\ r\la10^4$.  Therefore, in contrast
to the hot settling flow, thermal conduction in the SLE case is not
strong enough to modify the equilibrium structure of the flow.  We may
thus directly apply the stability criterion (\ref{stab-T}).  We then
find that the critical $k_{\rm crit}$ below which the thermal mode is
unstable is given by \beq k_{\rm crit}R \ga 180
\left(\vartheta_{-1}\dot m_{-2}\alpha_{-1}\right)^{1/2}r_2^{1/4}
\label{stab-SLE}
\eeq Since we find that $k_{\rm crit}R$ is very large, we conclude
that the SLE solution is thermally unstable even in the presence of
thermal conduction.  This is in contrast to the hot settling flow
where conduction has quite a dramatic effect.  The reason for the
difference is that the SLE solution is cooler and more collisional and
therefore has a significantly lower coefficient of thermal conduction.

\section{Conclusions}
\label{S:CONCL}

In this paper we have investigated how thermal conduction affects the
stability properties of two hot, cooling-dominated, rotating accretion
flows: the hot settling flow \citep{MN00} and the SLE
slim disk \citep{SLE76}.  Without thermal
transport, both flows are thermally unstable since they balance
viscous heating with free-free bremsstrahlung cooling \citep{P78}.
Thermal conduction smears out temperature perturbations and can in
principle suppress the instability. The stabilization effect is
proportional to $k^2 R^2$, where $k$ is the wave-vector of a
perturbation and $R$ is the radius, and is thus strongest for
small-scale perturbations, $kR\gg 1$.

To study the stability of the two rotating flows, we have used the
shearing sheet approximation.  We obtain the stability criterion
(\ref{stab-T}), which gives the critical wave-vector, $k_{\rm crit}$,
of a perturbation such that thermal modes with a larger spatial scale
($k<k_{\rm crit}$) are unstable.  We claim that an accretion flow is
thermally stable if $k_{\rm crit}R\la1$ and unstable if $k_{\rm
crit}R\gg1$.  Technically, for perturbations with $k\sim 1/R$, the
effects of geometry and curvature must be properly taken into
account. This requires a global analysis, which we do not attempt.

Because of the very high (nearly virial) temperature of the gas in the
hot settling flow, the particle mean-free-path is much larger than the
temperature gradient scale (which is comparable to $R$).  In this
regime, the Spitzer theory of thermal conduction cannot be applied.
Instead, one should consider collisionless transport theory including
the effects of tangled (but dynamically unimportant) magnetic fields.
The thermal flux is then given by equation (\ref{q-cond}).  We find
that this flux is energetically very important in the hot settling
flow and modifies even the equilibrium flow (see Appendix
\ref{A:SETTL}).  We allow for the modification and then analyse the
thermal stability of the resulting flow.  We obtain the stability
criterion (\ref{kR-stab-ss}), which indicates that modes with $kR\ga1$
are stable. This suggests that the hot settling flow is thermally
stable in the presence of thermal conduction.

The SLE solution is cooler than the hot settling flow, and particle
collisions are more frequent. The thermal flux in this flow is
described by collisional transport theory suitably modified for the
presence of tangled magnetic fields.  We find that thermal transport
in the SLE flow is not very important and that the flow is thermally
unstable even when conduction is included (see eq \ref{stab-SLE}).

We have thus answered the question with which we began the paper
(\S1).  However, in the process we have come up with an unexpected
realization: conduction in hot dilute accretion flows may be so large
as to substantially modify the equilibrium flow.  This raises three
interesting issues.

First, in the basic hot settling flow without conduction, it has been
shown that the energy source for viscous heating of the gas is
ultimately derived from the rotation of the central star 
\citep{MN00}.  If, as we have shown here, conductive heating is more
important than viscous heating, then one could ask where the energy
for the conductive flux comes from.  Clearly, the energy must be
derived from accretion.  A complete global solution, including
conduction, should be able to trace the flow of energy from the hot
inner regions of the accretion flow, where gravitational energy is
released, to the hot outer regions, where the gas is approximately
described by the self-similar settling flow solution.

Second, if particle transport can have such a strong influence on
energy balance, what about angular momentum balance?  Particles that
move from one radius to another will certainly have some effect on the
angular momentum of the gas, but the exact details are uncertain when
particle trajectories are constrained by the magnetic field.  For
simplicity, we have ignored in this paper angular momentum transport
by the collisionless particles.  This effect should be included for
consistency when the nature of the interactions is understood.

Third, apart from the hot settling flow, two other hot accretion flows
with nearly virial temperature are known: the ADAF and the CDAF.  One
suspects that energy conduction (and angular momentum transport) by
collisionless particles might be important for these solutions as
well, and might modify their equilibrium structures.  We note that a
CDAF is very different from an ADAF because convection has a strong
effect and changes the radial structure of the flow drastically
\citep{NIA00,QG00}.  If we now add
another form of transport, there could well be additional effects.

\acknowledgements 

This work was supported in part by NASA grant NAG5-10780.

\newpage
\begin{appendix}

\section{Collisionless thermal conduction in tangled magnetic fields}
\label{A:COND}

In this Appendix we discuss the essence of the thermal transport theory 
in collisionless systems threaded by weak tangled magnetic fields.

In the collisionless regime, the mean-free-path $\lambda$ of an
electron is comparable to or larger than the temperature gradient
scale $\lambda\ga T_e/|\nabla T_e|$. The heat flux in this case is
$\sim\rho c_s^3$. In a collisionless magnetized plasma, electrons
stream freely along the field lines only. In the transverse direction,
they are tied to the field lines on the Larmor scale. In a tangled
field, however, electrons can jump from one field line to another and
thus conduct heat across the field. This is the regime of interest to
us.  The physics of this regime of conduction has been discussed by
\citet{RR78} and \citet{CC98}.

Let us consider a tangled magnetic field with a characteristic
correlation scale $l_B$.  We assume that the electron Larmor radius
$\rho_e$ in this field is much smaller than the mean free path
$\lambda$.  If $l$ is the path length measured along a field line, the
separation between two closely neighboring field lines grows as \beq
d(l)\sim d(0)e^{l/L_K},
\label{d(l)}
\eeq where $d(0)$ is their initial separation and $L_K$ is the
Kolmogorov-Lyapunov length which, in general, depends on the field
spectrum. Since there is only one characteristic scale in the problem,
$l_B$, we expect $L_K\sim l_B$. Let us assume that at an initial
instant, an electron drifted a distance $d(0)\sim\rho_e$ from its
initial field line in the perpendicular direction.  Once it starts
moving along a new field line, its new path diverges from its initial
path according to equation (\ref{d(l)}). After the particle travels
the distance \beq L_{RR}\sim l_B\ln(l_B/\rho_e)
\label{L_RR}
\eeq measured along the field line, the separation becomes of order
$l_B$.  Any subsequent motion of the electron then becomes
uncorrelated with its initial field line. The distance $L_{RR}$ is
called the Rechester-Rosenbluth length.

If $v_T$ is the characteristic electron thermal speed, then from
dimensional considerations it follows that the thermal flux is
described by equation (\ref{q-grad}) with an effective conductivity
$\kappa_B\sim nk_Bv_T l_B$. A more rigorous analysis \citep{CC98}
shows that for large-scale perturbations, $kl_B\la1$, one can take the
ensemble average.  Then the thermal conductivity becomes isotropic
(assuming isotropic magnetic turbulence) and reads as \beq
\kappa_B\simeq nk_Bv_Tl_B\,\vartheta\,\frac{{\rm
min}\left(\lambda,L_{RR}\right)}{L_{RR}},
\label{kappa_B}
\eeq where the function ${\rm min}(a,b)$ denotes the smaller of $a$
and $b$.  The quantity $\vartheta\le1$ takes into account the fact
that only a fraction of particles (those which are not trapped by
magnetic mirrors) can move far enough to transport heat; the rest will
remain trapped in their local magnetic wells and be unable to
transport energy beyond a distance $\sim l_B$. We calculate
$\vartheta$ below.

There are three substantially different regimes of thermal conduction.
First, in the collisionless limit, $\lambda\gg L_{RR}$, we have
$\kappa_B\sim nk_Bv_Tl_B\vartheta$, i.e., $l_B$ plays the role of an
effective mean-free-path and only non-trapped particles contribute.
Second, in the semi-collisionless limit, $\rho_e\ll \lambda< L_{RR}$,
we have $\kappa_B\sim\kappa_{\rm Sp}\vartheta l_B/L_{RR}$.  That is,
the standard Spitzer collisional transport is affected by the presence
of tangled fields: the mean-free-path is set by collisions but most of
the particles are still trapped in magnetic mirrors.  Third, in the
strongly collisional limit: $\lambda\ll\rho_e$, the particles do not
move along the field lines, magnetic mirroring is inefficient and,
thus, $\kappa_B$ approaches the classical Spitzer value.  Note also
that in the first and second cases, the heat flux is anisotropic for
small-scale perturbations, $kl_B\gg1$, because the magnetic field is
locally ordered. The flux is given by equation (\ref{q-grad}) with
$\kappa=\kappa_B$ across the field lines, and by equation
(\ref{q-sat}) or equation (\ref{q-grad}) with $\kappa=\kappa_{\rm
Sp}$, depending on plasma collisionality, along the field lines.

Now we estimate the fraction of particles, $\vartheta$, which can pass
through magnetic mirrors. Let us consider a particle which moves along
a field line from a region ``1'' with a weak field to a region ``2''
with a stronger field. If the field gradient scale is much larger than
the Larmor radius (which is the case for most astrophysical systems),
the particle's magnetic moment $\mu=mv_\bot^2/2B$ ($v_\bot$ is the
velocity component perpendicular to the local field) is conserved as
an adiabatic invariant. That is $mv_{\bot1}^2/2B_1=mv_{\bot2}^2/2B_2$.
The energy of the particle is also constant. At the point where the
particle is reflected its parallel velocity vanishes, therefore
$v_{\bot2}^2=v_{\bot1}^2+v_{\|1}^2$. Combining these two equations we
obtain that particles are reflected from the magnetic mirror if their
pitch angles are greater than the minimal pitch angle, defined as:
\beq \sin^2\theta_m=B_{\rm min}/B_{\rm max}.  \eeq The particles with
$\theta<\theta_m$, which can pass though the mirror, form a ``loss
cone'', the solid angle of which is
$$
\Omega_{lc}=2\pi(1-\cos\theta_m).
$$
There is a second loss cone in the opposite direction along the 
field line, due to symmetry. Assuming an isotropic distribution of particle
velocities (in the weak field region), we estimate the fraction
of free streaming particles as
\beq
\vartheta=2\Omega_{lc}/4\pi
=1-\left(1-B_{\rm min}/B_{\rm max}\right)^{1/2}\sim0.3,
\eeq
where we assumed that the amplitude of magnetic turbulent fluctuations 
in an accretion flow is typically $\delta B\sim\langle B\rangle$,
i.e., $B_{\rm max}\sim2B_{\rm min}$.
\newpage

In hot accretion flows, the Coulomb mean-free-path is often much
larger than the size of the flow, so that $\lambda\ga L_{RR}$. It is
convenient then to write the conductive heat flux as in equation
(\ref{q-cond}), 
\beq q_{\rm cond}=-\alpha_c\frac{\rho c_s^2}{\Omega_K}\frac{dc_s^2}{dx}, 
\qquad{\rm where}\qquad
\alpha_c=\sum_{i=e,p}\kappa_B\frac{m_i\Omega_K}{k_B\rho c_s^2}\,
\frac{dc_{si}^2/dx}{dc_s^2/dx} \simeq2^{1/2}\frac{l_B}{H}\,\vartheta F(e,p) 
\eeq 
is the coefficient of thermal conduction, $i=e,p$ because
both species contribute to heat transport, and $F(e,p)\equiv{\rm
max}\left[F(e),F(p)\right]$ properly corrects for the two-temperature
flow, and where $H/R\simeq c_s/v_{\rm ff}=c_s/(\sqrt2\Omega_KR)$ is of
order unity in hot flows, $H$ is the accretion disk scale height and
$v_{\rm ff}$ is the free-fall speed. We also used here that
$v_{Te,p}\simeq c_{se,p}$. Thermal conduction is dominated by
electrons in a one-temperature flow because they are lighter than
protons. In a two-temperature flow, however, the protons may be much
hotter than the electrons and, hence, will dominate the conduction.
We write the relative contributions of electrons and protons to
thermal transport in two- and one-temperature zones:
\beq\begin{array}{rclrcl} F(e)_{2T}&=&\displaystyle
\frac{m_e}{m_p}\frac{c_{se}}{c_s}\frac{dc_{se}^2/dx}{dc_s^2/dx}, &
\qquad
F(e)_{1T}&=&\displaystyle\frac{1}{2\sqrt2}\sqrt{\frac{m_e}{m_p}}\simeq15.2,\\
F(p)_{2T}&=&1, & \qquad
F(p)_{1T}&=&\displaystyle\frac{1}{2\sqrt2}\simeq0.35,
\end{array}\label{F}
\eeq where $c_s^2=c_{sp}^2+(m_e/m_p)c_{se}^2=nk_B(T_p+T_e)$. We have
made use of the fact that in a two-temperature zone ($T_p\gg T_e$)
$c_s^2\simeq c_{sp}^2$ and in a one-temperature zone ($T_p=T_e$)
$c_s^2\simeq 2c_{sp}^2$.

\section{Properties of the hot settling flow onto a neutron star}
\label{A:SETTL}

The first analytical solution for a hot accretion flow settling onto a
rapidly rotating NS was found by \citet{MN00}. We briefly present here
some relevant results and derive some new results. The hot settling
flow is geometrically thick with the vertical thickness being
comparable to the local radius of the flow, $H\sim R$. The flow is hot
with the protons being at nearly the virial temperature. The flow
rotates with a sub-Keplerian angular velocity. The flow is optically
thin and is cooled by free-free emission. The flow is powered by the
rotational energy of the central star; hence the flow parameters (all
but the radial velocity) are independent of the mass accretion
rate. As in the case of an ADAF, the mass accretion rate must not
exceed a critical value which is of order a few percent of the
Eddington value; otherwise, the cooling is too strong and a thin
Shakura-Sunyaev disk forms.

The settling flow consists of two zones: an inner two-temperature zone
for $r\la10^{2.5}$ ($r=R/R_S$, where $R_S=2.95\times10^5m\textrm{ cm}$
is the Schwarzchild radius), and an outer one-temperature zone for
$r\ga 10^{2.5}$. Each zone is described by a self-similar solution. The
two- and one-temperature solutions read: 
\beq\begin{array}{rclrcl}
\rho_{2T}&\simeq&8.9\times10^{-3}m^{-1}\alpha s^2r^{-2}\textrm{g/cm}^3, 
&\qquad 
\rho_{1T}&\simeq&4.9\times10^{-2}m^{-1}\alpha s^2r^{-2}\textrm{ g/cm}^3,\\ 
\theta_{p,2T}&=&(1/6)r^{-1}, 
&\qquad \theta_{p,1T}&=&\theta_{p,2T}/2,\\ 
\theta_{e,2T}&\simeq&11r^{-1/2},
&\qquad \theta_{e,1T}&\simeq&153r^{-1},\\ 
\Omega_{2T}&\simeq& 7.2\times10^4m^{-1}sr^{-3/2}\textrm{ rad/s}, 
&\qquad \Omega_{1T}&=&\Omega_{2T},\\ 
v_{2T}&\simeq&1.4\times10^8\dot m\alpha^{-1}s^{-2}r^0\textrm{ cm/s}, 
&\qquad v_{1T}&\simeq&2.6\times10^7\dot m\alpha^{-1}s^{-2}r^0\textrm{ cm/s},
\label{ss}
\end{array}\eeq
where $\theta_p=k_BT_p/m_pc^2$ and similarly for the electrons,
$m=M_{NS}/M_{\sun}$,\ $\dot m=\dot M/\dot M_{\rm Edd}$ is the mass
accretion rate in Eddington units, $\dot M_{\rm
Edd}=1.4\times10^{18}m\textrm{ g/s}$,\ $\alpha$ is the viscosity
parameter, and $s=\Omega_{NS}/\Omega_{K}(R_{NS})$ is the NS spin in
units of the Keplerian angular velocity at the NS radius.  Note that,
except for $v$, none of the quantities depend on the accretion rate
$\dot m$. The one-temperature self-similar solution is valid from
$r\sim10^{2.5}$ to approximately \beq
r_{ss}\sim\left(74\alpha^2s^2/\dot m\right)^2 \sim
5.5\times10^3\alpha_{-1}^4 s_{-1}^4 \dot m_{-4}^{-2}.
\label{r-ss}
\eeq
Beyond this radius the solution is not self-similar.

The above self-similar solutions were obtained without taking into
account thermal conductivity of the gas. As we will now see, thermal
conduction in a hot settling accretion flow is very large, so that the
thermal flux is energetically important and modifies the structure of
the equlibrium settling flow itself. 

To be consistent with our simple analysis of the thermal instability
in \S\ref{S:TI}, it is sufficient to consider the one-temperature
regime only.  The hot settling solution is obtained from the condition
that the total local heating rate is equal to the total cooling rate.
Including the contribution from the divergence of the conductive flux,
the condition now reads \beq \vec\nabla\cdot\vec q_{\rm
cond}\equiv-\frac{1}{R^2}\frac{\partial}{\partial R}R^2
\frac{\alpha_c\rho c_s^2}{\Omega_K}\frac{\partial c_s^2}{\partial
R}=Q^++Q^-,
\label{energy-cond}
\eeq where $q_{\rm cond},\ Q^-$, and $Q^+$ are given by equations
(\ref{q-cond}), (\ref{heating}).  Since the additional heat flux term
has exactly the same radial dependence as the other two terms, the
power-law scalings given by equations (\ref{ss}) remain valid, but
with their pre-factors changed. Let us write the new solution as
$\rho=\hat\rho r^{-2},\ \theta_p=\hat\theta_p r^{-1}$, etc..  With
these definitions, the energy balance equation (\ref{energy-cond})
takes the form: \beq \frac{9}{4}\hat\rho
c^2\hat\theta_p\hat\Omega_Kr^{-9/2} \left(\alpha
s^2+\frac{4}{3}\hat\theta_p\alpha_c\right) ={\cal
C}\hat\rho^2c\hat\theta_p^{1/2}r^{-9/2}.  \eeq Clearly the effect of
thermal conduction is to to re-define the quantity
$\alpha s^2$ to \beq \alpha
s^2\to\alpha s^2+\frac{2}{9}\alpha_c,
\label{sub}
\eeq where we have used $\hat\theta_p=1/6$.  With this re-definition,
we may continue to use equations (\ref{ss}) for the equilibrium flow.

Next, we estimate the typical value of the thermal conduction
coefficient in the hot settling flow. From equations (\ref{alpha_c})
and (\ref{F}) we have \beq \alpha_c=2^{-1/2}\frac{R}{H}\,\xi\vartheta
F(e,p) \simeq10^{-2}\xi_{-1}\vartheta_{-1}F(e,p).
\label{alpha-c1}
\eeq From (\ref{F}) and (\ref{ss}) we obtain
$F(e)_{2T}\simeq0.15\,r^{3/4}\la10.9$ because the two-temperature zone
exists for $r\la10^{2.5}$. That is, in the two-temperature zone
$F(e,p)$ changes from $\sim 1$ in the inner parts to $\sim10$ in the
outer parts and then smoothly approaches the value of $\sim 15$ in the
one-temperature zone. Thus, the overall quantity $F(e,p)\equiv\textrm{
max}[F(e),F(p)]$ lies in the range $1\le F(e,p)\la15$ for the entire
flow. Moreover, we note that $F(e)\la F(p)$ only for small radii $r\la
13$, where the self-similar settling solution is not terribly accurate
because of Comptonization. Therefore, we see that thermal conduction
is dominated by the electrons over nearly the entire accretion flow.

Finally, we demonstrate that (\ref{kappa_B-am}) is the relevant limit
of equation (\ref{kappa_B}) and, hence, that the above estimate for
the thermoconduction coefficient is accurate. For this, we must
demonstrate that $\lambda\ga L_{RR}$ in the settling accretion flow
(see equation \ref{kappa_B}).

From the estimates above we have $\alpha
s^2+(2/9)\alpha_c\simeq2.2\times10^{-3}\xi_{-1}\vartheta_{-1}F(e,p)$.
From equation (\ref{mfp}), the electron mean-free-path,
$\lambda\sim0.57\theta_e^2/\rho$, normalized by the local radius of
the flow, is estimated in both zones to be \beq \lambda_{2T}/R \simeq
12\left[\xi_{-1}\vartheta_{-1}F(e,p)\right]^{-1}, \qquad
\lambda_{1T}/R \simeq
17\left[\xi_{-1}\vartheta_{-1}F(e,p)\,r_3\right]^{-1}.
\label{lambda2}\eeq
The proton mean free path is comparable to or larger than the above
estimate. To calculate $L_{RR}$ the Larmor radius of the electrons 
is needed. It may be estimated as follows.
The magnetic pressure $P_m=P_{\rm gas}/\beta$ yields
$$
B\simeq\left(8\pi\rho c_s^2/\beta\right)^{1/2}.
$$
Since $c_s^2=c_{sp}^2+(m_e/m_p)c_{se}^2\sim c_{sp}^2$ to within a factor of 
two and the thermal velosity of particles is $v_{Te}\simeq c_{se}$, we obtain
\beq
\rho_L\simeq v_{Te}/\Omega_B
\simeq1.1\times10^{-8}\left(\frac{\beta\theta_e}{\rho\theta_p}\right)^{1/2},
\label{Larmor}
\eeq
where $\Omega_B=eB/m_ec$ is the electron cycloron frequency.
For a typical $\beta\simeq10$ (see e.g., \citealp{M00,QG99}), 
the normalized electron Larmor radii 
in the two- and one-temperature zones are:
\beq
\rho_{L,2T}/R \simeq 2.2\times10^{-10}m^{-1/2}r^{1/4}
\left[\xi_{-1}\vartheta_{-1}F(e,p)\right]^{-1/2}, \qquad
\rho_{L,1T}/R \simeq 5.1\times10^{-10}m^{-1/2}
\left[\xi_{-1}\vartheta_{-1}F(e,p)\right]^{-1/2}.
\label{rhoL2}\eeq
Since the two quantities differ at most by a factor of order unity, we
use $\rho_{L}/R\sim5\times10^{-10}$ as a representative value.  Now we
can estimate the Rechester-Rosenbluth length from equation
(\ref{L_RR}), \beq L_{RR}/R\simeq\xi\ln\left[\xi/(\rho_L/R)\right]
\simeq10^{-1}\xi_{-1}(19+\ln\xi_{-1})\sim1.9\xi_{-1}.  \eeq Finally,
we have \beq
\lambda_{2T}/L_{RR}\simeq6.3\left[\xi_{-1}^2\vartheta_{-1}F(e,p)\right]^{-1},
\qquad
\lambda_{1T}/L_{RR}\simeq8.9\left[\xi_{-1}^2\vartheta_{-1}F(e,p)\right]^{-1}
\eeq We see that in the two-temperature zone, the mean free path is
large, $\lambda\ga L_{RR}$, so that equations (\ref{kappa_B-am}) and
(\ref{kappa_B}) are justified. In the one-temperature zone, the result
is marginal: $\lambda\sim L_{RR}$.  However, considering that (i) the
transition between the regimes with $\lambda\ll L_{RR}$ and
$\lambda\gg L_{RR}$ is not sharp in reality, (ii) the ``boundary''
itself, $\lambda \sim L_{RR}$, is uncertain to within a numerical
factor of a few, and (iii) the one-temperature zone is rather small
[c.f., equation (\ref{r-ss})], we believe that we may use equation
(\ref{kappa_B-am}) throughout the flow.  The error will not be greater
than a factor of order unity.

\section{Properties of the SLE solution}
\label{A:SLE}

The SLE slim disk solution was the first hot accretion solution found.
The SLE solution is optically thin. It is cooling-dominated via
free-free emission (the original paper by SLE considered Compton
cooling).  As in other hot solutions, there are both two- and
one-temperature zones in this flow. In the two-temperature zone, the
normalized electron and proton temperatures are related as follows:
\beq
\theta_e=\left(\frac{\pi\ln\Lambda}{12\alpha_f}\right)^{1/2}\theta_p^{1/2}
\simeq23\theta_p^{1/2},
\label{SLEtheta}
\eeq
where $\alpha_f$ is the fine structure constant and $\ln\Lambda=15$.
This equation is easily obtained from the equality of the electron
cooling rate via Bremsstrahlung and the energy transfer rate from
the protons to the electrons via Coulomb collisions, $Q^-=Q_{\rm Coul}$;
the expressions for these rates are given in \citet{MN00}. In the 
one-temperature zone, $T_p=T_e$, i.e., $\theta_e=(m_p/m_e)\theta_p$.
Following \citet{SLE76}, we obtain the following solutions:
\beq\begin{array}{rclrcl}
c_{s,2T}&\simeq&6.5\times10^9\dot m^{1/4}\alpha^{-1/2}r^{-3/8}\textrm{ cm/s}, 
&\qquad
c_{s,1T}&\simeq&1.9\times10^9\dot m^{1/4}\alpha^{-1/2}r^{-3/8}\textrm{ cm/s}, \\
\rho_{2T}&\simeq&2.9\times10^{-3}m^{-1}\dot m^{1/4}\alpha^{1/2}r^{-15/8}
\textrm{ g/cm}^3, &\qquad
\rho_{1T}&\simeq&1.2\times10^{-1}m^{-1}\dot m^{1/4}\alpha^{1/2}r^{-15/8}
\textrm{ g/cm}^3, \\
(H/R)_{2T}&\simeq&3.1\times10^{-1}\dot m^{1/4}\alpha^{-1/2}r^{1/8},  &\qquad
(H/R)_{1T}&\simeq&2.2\times10^{-1}\dot m^{1/4}\alpha^{-1/2}r^{1/8}, 
\label{SLE-sol}
\end{array}\eeq

From the above equations, the dimensionless temperature in the two-temperature
zone ($r\la10^2$) is $\theta_p=(c_s/c)^2\simeq0.05\ r^{-3/4}$ which is 
significantly smaller than the temperature of the hot settling flow
in which $\theta_p\simeq0.2 r^{-1}$. In the one-temperature zone,
$theta_p\simeq0.004\ r^{-3/4}$ is again smaller that the temperature of
the hot settling flow, $\theta_p\simeq0.1\ r^{-1}$, up to a very large
radius $r\sim4\times10^{6}$. We see that the temperature of the gas in 
the SLE disk is smaller than in the hot settling flow. Therefore, we
expect that collisions are more frequent and they may change the
thermal properties of the gas.  Indeed, we denonstrate now that
$\lambda\la L_{RR}$ in the SLE disk, that is, thermal conduction is in
the semi-collisionless regime (see Appendix \ref{A:COND}).

From equations (\ref{mfp}) with $T=c_s^2m_p/k_B$ and (\ref{SLEtheta}),
we obtain the normalized collisional mean-free-paths \beq
\lambda_{2T,p}/R\simeq2.9\dot m^{3/4}_{-2}
\alpha_{-1}^{-5/2}r_{2}^{-5/8},\qquad \lambda_{2T,e}/R\simeq0.3\dot
m^{1/4}_{-2} \alpha_{-1}^{-3/2}r_{2}^{1/8},\qquad
\lambda_{1T}/R\simeq2.2\times10^{-2}\dot m^{3/4}_{-2}
\alpha_{-1}^{-5/2}r_{2}^{-5/8}.
\label{lambdaSLE2}\eeq
The normalized electron Larmor radius follows from equation (\ref{Larmor}):
\beq
\rho_{L,2T}/R\simeq1.3\times10^{-10}m^{-1/2}
\dot m^{-1/4}_{-2}r^{1/8}_{2}, \qquad
\rho_{L,1T}/R\simeq3.6\times10^{-11}m^{-1/2}
\dot m_{-2}^{-1/8}\alpha_{-1}^{-1/4}r_{2}^{-1/16},
\label{rhoLSLE2}\eeq
where we take $r\sim100$ as a characteristic radius at which a
one-temperature flow becomes two-temperature. Since
$\rho_L\propto(Tm)^{1/2}$, the proton Larmor radius is
$\sqrt{m_p/m_e}\sim43$ times larger.  We estimate $L_{RR}$ from
equation (\ref{L_RR}) as \beq
L_{RR}/R\sim-\xi\ln(\rho_L/R)\sim2.5\xi_{-1} \eeq for the electrons
and it is slightly less for the protons.  Comparing this with
(\ref{lambdaSLE2}), we see that $\lambda\ll L_{RR}$ in that part of
the disk where the electron thermo-conduction dominates.  Only in the
inner parts of the disk, where the proton conductivity becomes large,
do we have $\lambda\ga L_{RR}$.

Finally, we calculate the thermal conductivity of the gas in the SLE disk.
It is determined by equation (\ref{kappa_B}) with $\lambda < L_{RR}$:
\beq
\kappa_B=\kappa_{\rm Sp}\vartheta(l_B/L_{RR})
=\kappa_{\rm Sp}\vartheta/|\ln(\rho_L/R)|
\simeq4\times10^{-3}\vartheta_{-1}\kappa_{\rm Sp}\ll\kappa_{\rm Sp}.
\label{kappa_B-SLE}
\eeq This thermoconductivity is about two orders of magnitude smaller
than the Spitzer value. As we discussed above, the protons may
significantly contribute to the thermal conduction flux and may even
dominate the electron contribution in the inner parts of the disk.
The proton contribution, which is $\kappa_{{\rm Sp}p}=\sqrt{m_e/m_p}
\kappa_{{\rm Sp}e}(T_p/T_e)^{5/2}$ (see equations [\ref{kappa-Sp}],
[\ref{mfp}]), becomes dominant in the two-temperature zone with
$T_p\gg T_e$. In one-temperature zone, the electron contribution
dominates, since $\kappa_{{\rm Sp}p}\simeq(1/43)\kappa_{{\rm Sp}e}$.
Using the self-similar SLE solution (\ref{SLE-sol}), we calculate the
{\em nominal} Spitzer thermal conductivity in the SLE disk \beq
\kappa_{{\rm Sp},2T}\simeq2.7\times10^{21} \dot
m^{5/4}\alpha^{-5/2}r^{-15/8},\qquad \kappa_{{\rm
Sp},1T}\simeq2.5\times10^{20} \dot m^{5/4}\alpha^{-5/2}r^{-15/8}
\label{kappa_Sp-SLE}
\eeq in CGS units. The real value of $\kappa$ is less than the Spitzer
value by the factor given in (\ref{kappa_B-SLE}).

\end{appendix}

\newpage
\figcaption[dispersion.eps]{Real part (left panel) and imaginary part
(right panel) of the frequencies of three modes.  The curves labeled 1
and 2 refer to the two acoustic modes, and the curves labeled 3 refer
to the thermal mode.  The dashed curves correspond to the dispersion
relation (\ref{disp0}), which does not include thermal conduction.
The following parameter values were used: $\kappa_{\rm
epi}/\Omega_K=s= \sqrt{0.5}$, $\tau_{\rm cool,ff}=10\Omega_K^{-1}$,
$c_s=\Omega_K R$, $\gamma=5/3$.  The mode frequencies are normalized
by the Keplerian frequency.  The solid curves correspond to the
dispersion relation (\ref{disp1}), which includes thermal conduction.
Here, $\tau_{\rm cond}=\tau_{\rm cool,ff}$, and the other parameters
are the same as before.
\label{f:disp} }

\bigskip
\plottwo{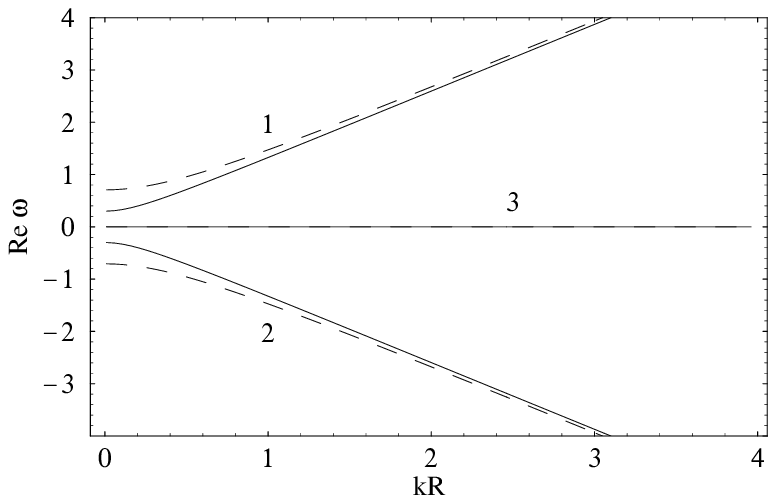}{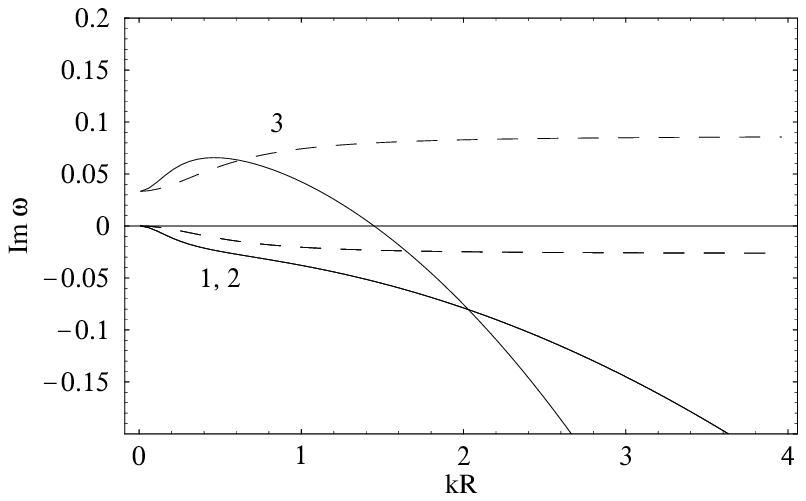}

\end{document}